\documentclass[preprint,5p,times,authoryear]{elsarticle}
\usepackage{amsmath,pifont,graphicx,amssymb,color,setspace,lineno}
\begin{document}
\begin{frontmatter}
\title{The cool and distant formation of Mars}
\author[elsi]{R.~Brasser\corref{cor}\fnref{fncrio}}
\cortext[cor]{Corresponding author}
\ead{brasser\_astro@yahoo.com}
\author[cu,has]{S. J. Mojzsis\corref{cor}\fnref{fncrio}}
\ead{mojzsis@colorado.edu}
\author[dun]{S. Matsumura\fnref{fn}}
\author[elsi]{S. Ida}
\address[elsi]{Earth Life Science Institute, Tokyo Institute of Technology, Meguro-ku, Tokyo 152-8550, Japan}
\address[cu]{Department of Geological Sciences, University of Colorado, UCB 399, 2200 Colorado Avenue, Boulder, Colorado 80309-0399, 
USA}
\address[has]{Institute for Geological and Geochemical Research, Research Center for Astronomy and Earth Sciences, Hungarian 
Academy of Sciences, 45 Buda\"{o}rsi Street, H-1112 Budapest, Hungary}
\fntext[fncrio]{Collaborative for Research in Origins (CRiO)}
\address[dun]{School of Science and Engineering, Division of Physics, Fulton Building, University of Dundee, Dundee 
DD1 4HN, UK}
\fntext[fn]{Dundee Fellow}
\begin{abstract}
With approximately one ninth of Earth's mass, Mars is widely considered to be a stranded planetary embryo that never became a 
fully-grown planet. A currently popular planet formation theory predicts that Mars formed near Earth and Venus and was subsequently 
scattered outwards to its present location. In such a scenario, the compositions of the three planets are expected to be similar to 
each other. However, bulk elemental and isotopic data for martian meteorites demonstrate that key aspects of Mars' composition are 
markedly different from that of Earth. This suggests that Mars formed outside of the terrestrial feeding zone during primary accretion. 
It is therefore probable that Mars always remained significantly farther from the Sun than Earth; its growth was stunted early and its 
mass remained relatively low. Here we identify a potential dynamical pathway that forms Mars in the asteroid belt and keeps it outside 
of Earth's accretion zone while at the same time accounting for strict age and compositional constraints, as well as mass differences. 
Our uncommon pathway (approximately 2\% probability) is based on the Grand Tack scenario of terrestrial planet formation, in which the 
radial migration by Jupiter gravitationally sculpts the planetesimal disc at Mars' current location. {\ We conclude that Mars' 
formation requires a specific dynamical pathway, while this is less valid for Earth and Venus. We further predict that} Mars' volatile 
budget {\ is most likely} different from Earth's and that Venus formed close enough to our planet that it is expected to have a {\ 
nearly identical} composition from common building blocks.
\end{abstract}
\end{frontmatter}
\section{Introduction}
\label{sec:int}
The formation of the terrestrial planets is a long-standing problem that is gradually being resolved. The past decade has witnessed 
important progress towards a unified model of terrestrial planet formation. From analysis of samples of the oldest-known rocks 
collected on Earth and the Moon, from lunar, martian and asteroidal meteorites, as well as remote sensing studies, we now have 
information on the nature and timing of formation of several worlds in our solar system through combined geochemical models, elemental 
and isotopic abundances, and geochronology. Analysis of martian meteorites show that it formed within $\sim$10 Myr of the start of the 
solar system \citep{D11}. The chemical and mechanical closure of Earth’s metallic core, as derived from the Hf-W chronometer, took 
place at least 20 Myr later than this (e.g. \citet{K09} and references therein). Adding these observations together leaves us with a 
general timeline for the formation of the terrestrial planets, and thus a foundation for computational models to explain their 
history.\\

In traditional dynamical models the terrestrial planets grow from a coagulation of planetesimals into protoplanets and subsequently 
evolve into a giant impact phase, during which the protoplanets collide with one other to give rise to the terrestrial worlds. 
Several variations of this scenario exist. {The most recent of these, dubbed `pebble accretion' (e.g. \citet{L15} and references 
therein), postulates that the terrestrial planets grow directly from the accretion of a swarm of centimetre-sized planetesimals termed 
pebbles; the outcomes of the pebble accretion model are presently an area of much active research. For this work, however, we shall 
make use of the popular and more established} {\it Grand Tack} model \citep{W11}.\\

Grand Tack relies on early gas-driven migration of Jupiter and Saturn to gravitationally sculpt the inner solid circum-solar disc 
down to $\sim$1~AU after which terrestrial planet formation proceeds from solids in an annulus ranging from roughly 0.7~AU to 1~AU. 
Grand Tack has booked some successes, such as its ability to reproduce the mass-orbit distribution of the terrestrial planets, the 
compositional gradient, and total mass of the asteroid belt \citep{W11}. Its predictions for the composition of the terrestrial 
planets, however, have not been widely explored. Through a combination of geochemical and N-body simulations we here report what the 
Grand Tack models predicts for the variation in the bulk compositions of the terrestrial planets, with particular focus on Mars. {\ 
The aim of this study is to constrain Mars' building blocks and whether these are identical to those of Earth's.}

\section{Isotopic heterogeneities}
Geochemical data from martian meteorites suggests that the overall composition of Mars is unlike that of the Earth (and Moon). Early 
investigations into Mars' bulk composition concluded that its primary constituents are a highly reduced component devoid of most 
volatiles, and more oxidised material that follows CI abundances \citep{WD88}; these are present in an approximately 2:1 ratio 
\citep{WD94}. These same studies concluded that Mars accreted homogenously, while Earth did not -- cf. \citet{D04}. Based on the 
analysis of the isotopic variations in O, Cr, Ti and Ni in various meteorites, terrestrial and martian rocks, this composition ratio 
was recently revised: Mars is a mixture of carbonaceous and non-carbonaceous material, with the former contributing only 9\%; for 
Earth the fraction is 24\% \citep{War11}. The link between isotopic anomalies and bulk composition is debated, but good cases 
{\ for such a link have been made in the past} \citep{War11,D14,D15}. Therefore, in this study, we will follow these works and 
assume that isotopic anomalies are correlated to differences in bulk compositions. \\

The three oxygen isotope system (expressed in the conventional $\Delta ^{17}{\rm O}=\delta ^{17}{\rm O}_{\rm VSMOW}-0.52\delta 
^{18}{\rm O}_{\rm VSMOW}$ notation; the $\delta$-notation denotes deviations in parts-per-thousand {\ where isotopic ratios of the 
same element are normalised to a standard}) permits discrimination between primordial nebular heterogeneities inherited during planet 
formation and mass-dependent planetary processes \citep{CM83}. Mars is defined by oxygen that is significantly enhanced in the minor 
isotope ($^{17}$O) with respect to terrestrial and lunar values \citep{IAF99,R11,Mit08,AG13,W15}, suggesting that the mixture of 
source components was different for Earth and Mars {\citep{WD88,WD94,L00,War11}}, which implies different source locations from 
within different compositional reservoirs of the solar nebula -- cf. \citet{FB16}.\\

This conclusion is lent weight by several recent isotope studies in meteorites and in terrestrial and martian samples. The terrestrial 
isotopic composition of $^{17}$O, $^{48}$Ca, $^{50}$Ti, $^{62}$Ni and $^{92}$Mo is best reproduced by a mixture of 90\% enstatite 
chondrite, 7\% ordinary chondrite and 2\% carbonaceous chondrites \citep{D14}. In contrast, for Mars a mixture of {\ 45\% enstatite 
chondrite and 55\% ordinary chondrite} can match its $^{17}$O, $^{50}$Ti, $^{54}$Cr, $^{62}$Ni and $^{92}$Mo values \citep{S99,TD14}, 
which is different from Earth's and thus hints at a formation region well away from that of Earth.\\

\begin{figure*}[ht!]
\includegraphics[width=90mm]{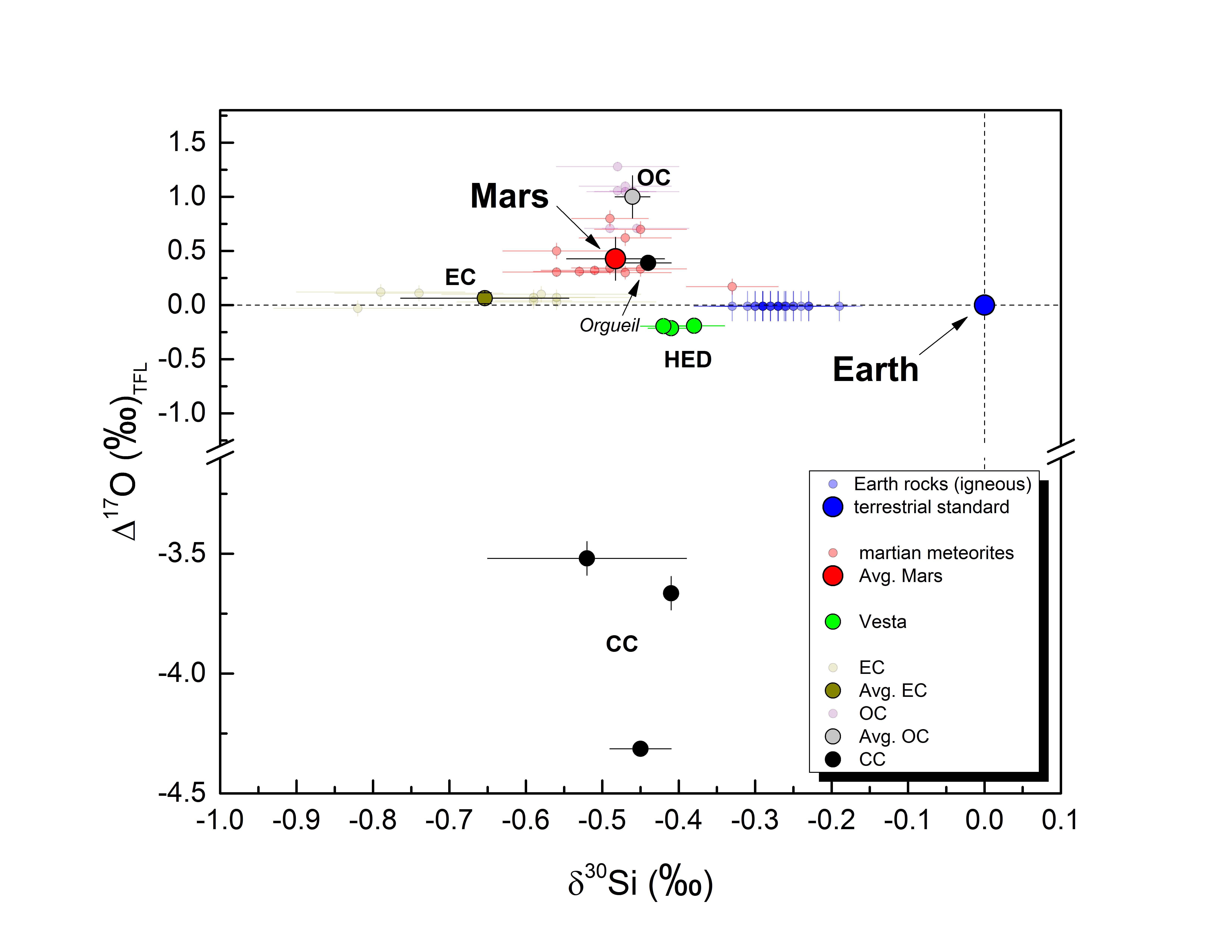}
\includegraphics[width=90mm]{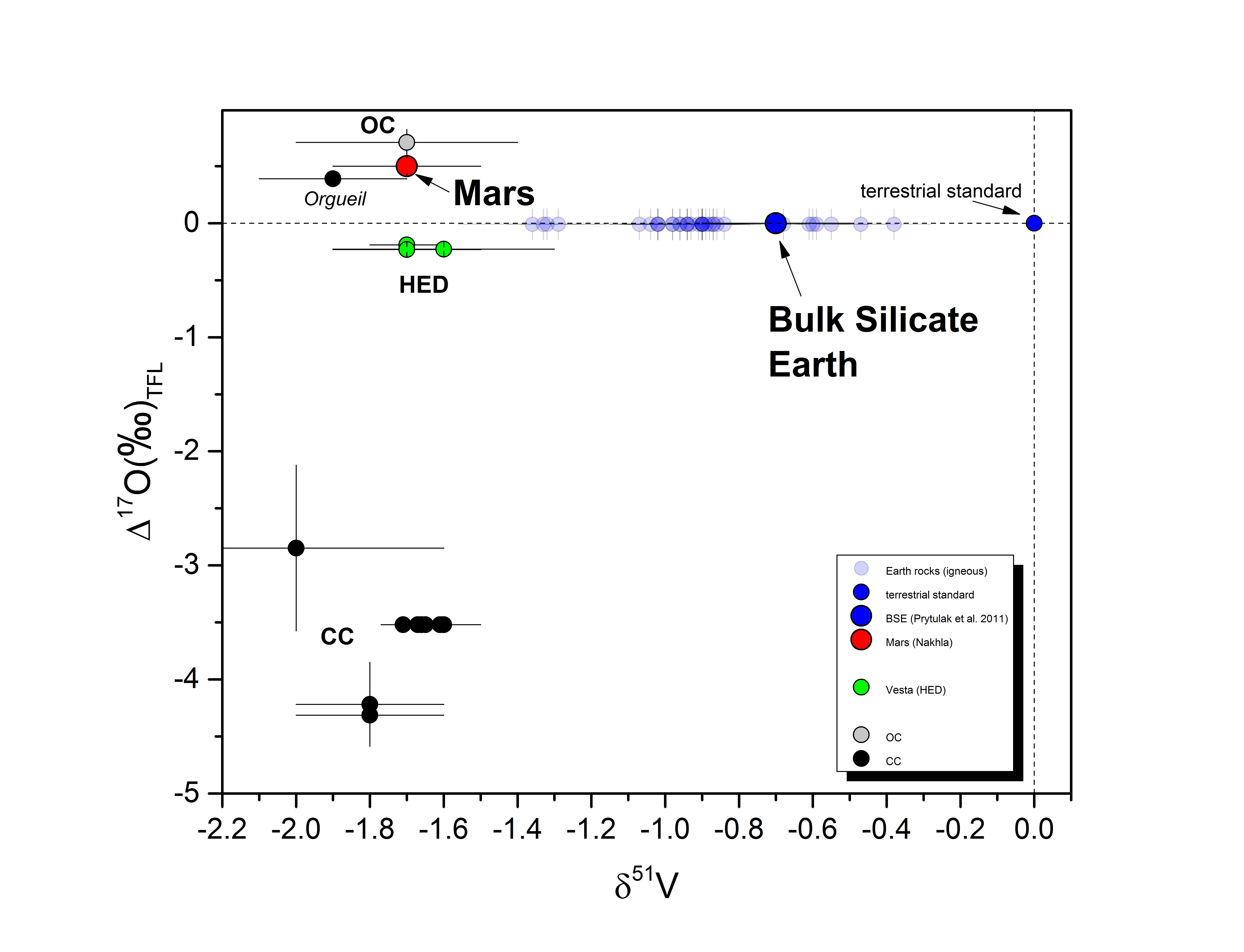}
\includegraphics[width=90mm]{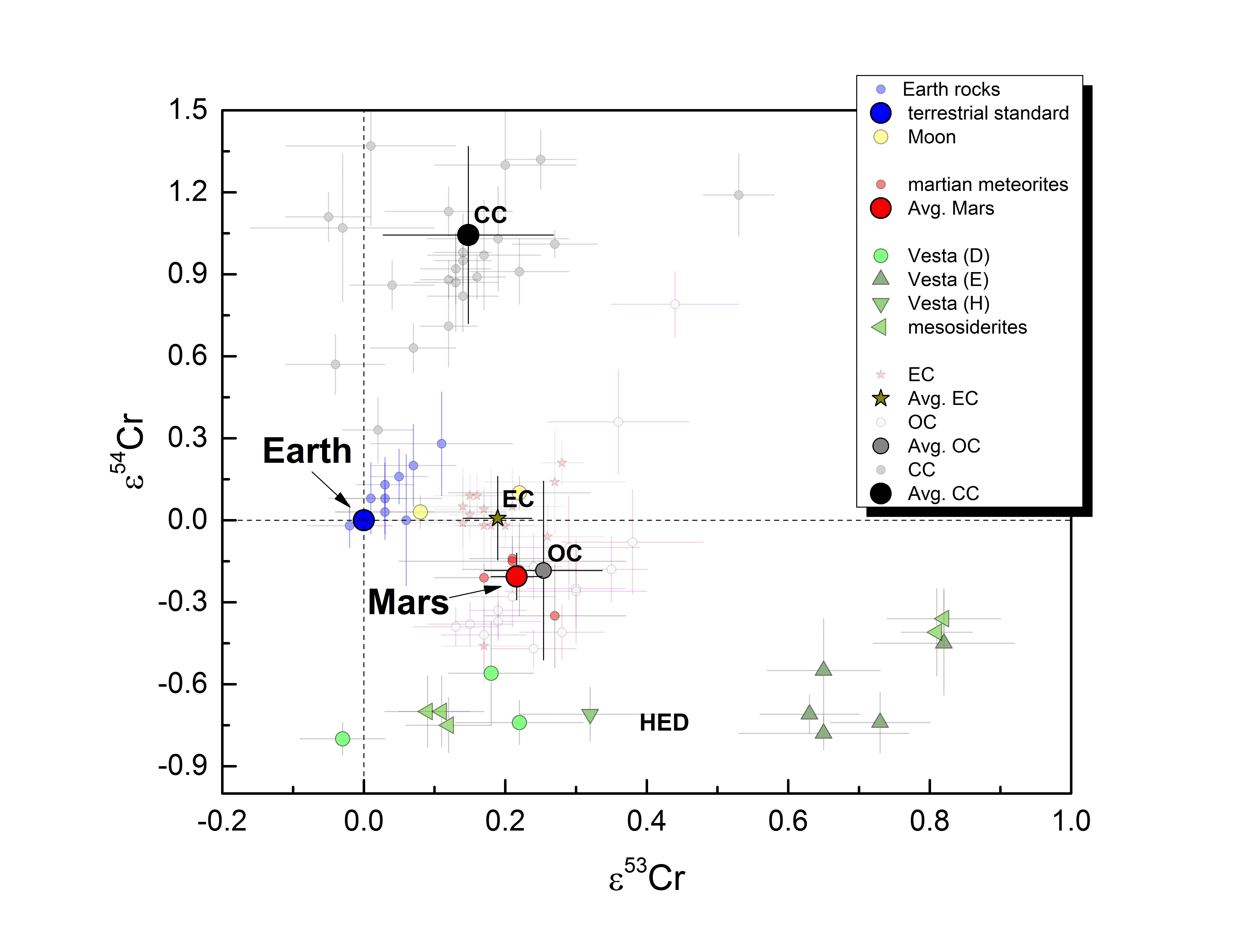}
\includegraphics[width=90mm]{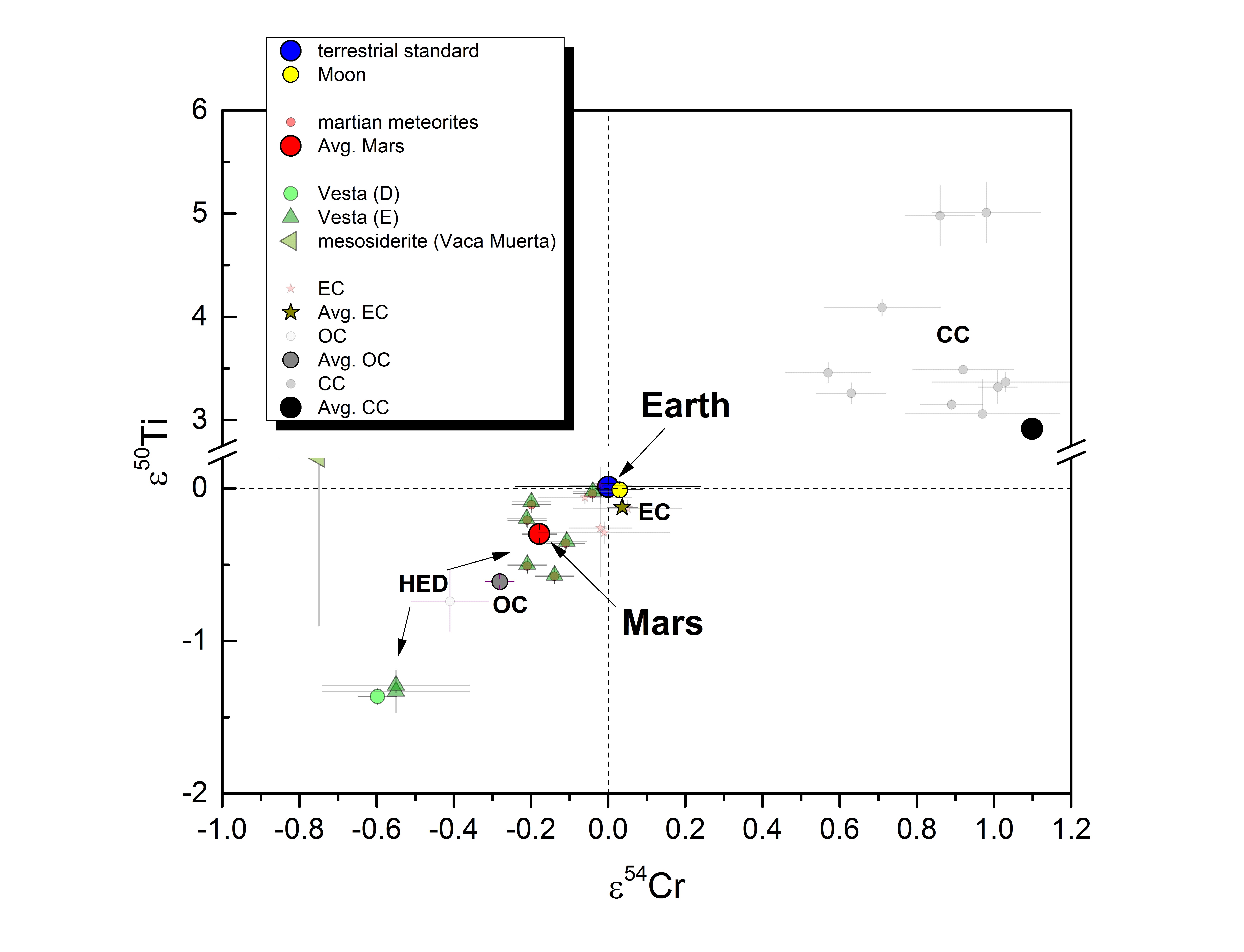}
\caption{Four-panel plot of the correlated comparative isotopic compositions of Earth (and Moon), Mars, Vesta (howardites, eucrites, 
diogenites, mesosiderites), EC (enstatite chondrites), OC (ordinary chondrites), and CC (carbonaceous chondrites). All data are 
normalised according to the terrestrial standard values. We only plot correlated data by sample, and not just averages, for each 
meteorite group. As such, differences that exist between different samples in, for example oxygen isotopes in the HEDs, become 
highlighted. Top-left: correlated oxygen vs. silicon. Top-right: correlated oxygen vs. vanadium. Bottom-left: correlated multiple 
chromium isotopes ($^{53}$Cr and $^{54}$Cr). Bottom-right: correlated titanium and chromium isotopes ($^{50}$Ti and $^{54}$Cr).}
\label{fig:isotopes}
\end{figure*}

Here we build upon these previous studies and highlight isotopic differences between Earth {(cf. \citet{J10})} and Mars and major 
meteorite groups. In Fig.\ref{fig:isotopes} we present comparative Si vs. $\Delta ^{17}$O (top-left panel), V vs. $\Delta ^{17}$O 
(top-right panel), multiple-Cr (bottom-left panel), and Ti vs. Cr (bottom-right panel) isotope data for Mars vs. Earth, and when 
compared to various meteorite groups for which data are available and correlative to the same sample. Correlated silicon and oxygen 
isotopes for Mars generally match the ordinary chondrites \citep{Georg07,P13a}, {\ indicating there is little to no silicon in Mars' 
core}, but Earth and Moon do not. The non-chondritic Si isotope composition of the Earth's mantle points to Si incorporation into the 
core \citep{Georg07}. {\ There is some debate in the literature regarding the source of Si fractionation. One argument revolves 
around nebular fractionation of SiO in fosterite at high temperature \citep{D15}, while another suggests Si is fractionated during 
impact-induced evaporation rather than core formation \citep{P14}.} Therefore, {\ it is possible} that the lower $\delta ^{30}$Si of 
Mars compared to Earth implies a more distant formation from the Sun, where temperatures were cooler. {\ If the latter mechanism 
dominates then Mars' lower escape velocity compared to Earth could be the source of its lighter Si isotopic composition.}\\

The correlated vanadium isotopic composition, expressed as $\delta ^{51}$V, vs. $\Delta ^{17}$O for Mars is 5$\sigma$ from terrestrial 
standards, which cannot be explained by metal-silicate partitioning \citep{N14}. Both $\varepsilon ^{50}$Ti and $\varepsilon ^{54}$Cr, 
as well as $\varepsilon ^{62}$Ni (the $\varepsilon$-notation denotes deviations in parts-per-ten thousand {\ normalised to 
another standardised isotopic ratio of the same element}), reflect the presence of planetary-scale nucleosynthetic anomalies 
\citep{War11,TD14}, while the origin of $\delta ^{51}$V variations is unknown.\\

The neutron-rich isotopes are suggested to have been implanted into the proto-solar disc by nearby supernovae \citep{Q10}, which is 
more effective at larger distances from the Sun, and could thus explain the $\varepsilon ^{54}$Cr vs $\varepsilon ^{62}$ Ni trend 
observed across various meteorite groups \citep{War11}. Studies have shown that, like in the O, Si and V systems cited above, major 
meteorite groups also possess $^{54}$Cr/$^{52}$Cr vs. $^{53}$Cr/$^{52}$Cr values that show clear differences from Earth 
\citep{T07,T08,Q10}. Of the Cr isotopes $\varepsilon^{53}$Cr is a tracer for volatility, and thus formation distance from the 
Sun, so that Mars' depletion in $\varepsilon ^{53}$ Cr relative to Earth hints at a cooler formation environment. Finally, the 
neutron-rich system, $^{50}$Ti/$^{47}$Ti (expressed as $\varepsilon ^{50}$Ti with respect to the terrestrial standard), contains 
anomalies comparable to the $\varepsilon ^{54}$Cr values for the same meteorites and components \citep{T09}. Hence, multiple lines 
of evidence show that different bulk elemental and isotopic makeups of Earth and Mars point to different accretionary histories and 
therefore source regions for the two planets.\\

In an alternative view presented by \citet{FB16}, Monte Carlo mixing models for a subset of isotopic systems can be devised that yield 
a singular mixture of various achondrite and chondritic components - some of which were likely comprised of 
compositionally-unconstrained differentiated planetesimals - that provide an Earth and Mars bulk composition. Such Monte Carlo 
mixing models ultimately formulate a compositional fit for both Earth and Mars out of 10$^{10}$ trials from 18 variables. 
{\ \citet{FB16} report that Earth and Mars are built from up to 93\% of the same material.} It is important to note, however, that 
the Monte Carlo mixing method neglects to take the dynamics of planetary formation into account. {\ The authors state that a 
scenario where Mars forms beyond 2~AU is inconsistent with the outcome of their experiments and instead prefer a formation model where 
Mars and Earth share the same feeding zone i.e. one in which they both accrete from very similar (well-mixed) materials}.\\

Our approach is instead to track compositions of the forming planets with dynamical simulations that yield Mars while 
simultaneously accounting for observational constraints. In the next section we argue for Mars' building blocks to predominantly 
consist of {\ a specific mixture of meteorite parent bodies, just like most of the other terrestrial planets \citep{War11}. We 
further argue from the isotope data {\ and our dynamical simulations} that bulk Mars is composed of different material than bulk 
Earth.} Therefore Mars most likely did not form in Earth's feeding zone, and conversely grew much farther from the Sun than the Earth. 
This conclusion may seem obvious given the present location of Mars in the Solar System, but recent dynamical models have all but 
invalidated it (see \citet{Morby12} for a review). Here we show that this idea is indeed consistent with recent numerical models of 
terrestrial planet formation, provided that Mars followed a specific dynamical path.

\section{N-body simulations and the Grand Tack}
In the classical accretion model the terrestrial planets remain more or less at their current positions, with some radial mixing 
caused by perturbations from the giant planets. In this model, however, the mass of Mars was always much higher than in reality 
\citep{R09}. A proposed solution initially confined all solid material to an annulus between the present positions of Venus and Earth 
at 0.7 AU and 1 AU, and with this setup the mass-distance relationship of the terrestrial planets was nicely reproduced in N-body 
simulations \citep{H09}. We refer to this model as the ‘annulus model’. A second relies on the terrestrial planets having formed from 
small pebbles as they spiralled towards the Sun \citep{L15}, the so-called pebble accretion model. Thus, the standard model has mostly 
been abandoned in favour of three alternatives that confine or deliver most of the solid mass within 1~AU: the Grand Tack \citep{W11} 
model, the annulus model \citep{H09} and pebble accretion \citep{L15}. All three can reproduce the large mass ratio between Mars and 
Earth. The last of these is still in its infancy, however, and the outcome of pebble accretion N-body simulations are sensitive to the 
underlying disc structure, leading to a great variety of outcomes \citep{I16}. For this reason we make use of the Grand Tack.\\

Briefly, Grand Tack relies on the early, gas-driven radial migration of Jupiter. After the gas giant formed it opened an annulus in 
the gas disc and began migrating towards the Sun \citep{LP86} because torques acting on the planet from said disc are imbalanced. At 
the same time, beyond Jupiter, Saturn slowly accreted its gaseous envelope. Once Saturn reached a critical mass of about 50 Earth 
masses it migrated Sunward too, then rapidly caught up with Jupiter and became trapped in the 2:3 mean motion resonance \citep{MS01}. 
This particular configuration of orbital spacing and mass ratio between the gas giants reversed the total torque on both planets, thus 
reversing their migration; the planets ‘tacked’. To ensure Mars did not grow beyond its current mass, the location of this ‘tack’ was 
set at 1.5~AU \citep{W11}. The Grand Tack scenario can further account for the mass-semimajor axis distribution of the terrestrial 
planets and the demography of the asteroid belt \citep{W11,DC14}.\\

The dynamical consequences of the Grand Tack model on the terrestrial planet region are two-fold: First, as Jupiter migrated towards 
the Sun, it strongly scattered all protoplanets and planetesimals in its path. This scattering placed about half of the solid mass of 
the disc (equivalent to about 1.8 Earth masses) from the inner solar system to farther than 5~AU from the Sun. Second, Jupiter acted 
as a snowplough that pushed some protoplanets and planetesimals towards the Sun. This shepherding of material mostly occurred within 
the 2:1 mean motion resonance with the planet. When Jupiter reversed its migration at 1.5~AU it caused a pileup of solid material 
inside of 1~AU at the location of the 2:1 resonance. Thus, almost all protoplanets and planetesimals that originally formed between 
1~AU and 1.5~AU were cleared by Jupiter, with a few pushed inside of 1~AU. Owing to the fact that Mars is currently at 1.5~AU from the 
Sun and formed farther than the Earth, it escaped Jupiter’s Grand Tack incursion somehow. We propose a scenario wherein proto-Mars 
formed in a few million years \citep{D11,TD14} farther than 1.5~AU from the Sun. The still-growing, but nearly-formed Mars 
subsequently migrated inwards to 1.5~AU by the dynamical interaction with Jupiter, other protoplanets and planetesimals in the disc. \\

To test our hypothesis we ran a high number of N-body simulations in the framework of the Grand Tack model with the tack at 1.5~AU and 
2~AU. All of these simulations and the data used here are discussed in our earlier analysis of the Grand Tack model \citep{B16}. We 
briefly summarise our initial conditions and methods here.\\

The initial conditions consist of a solid disc of sub-Mars sized planetary embryos and equal-mass planetesimals, with Jupiter and 
Saturn situated farther out. {\ The individual embryo masses are either 0.025, 0.05 or 0.08 $M_\oplus$. To account for the 
evolution of the gas in the disc during the time that the embryos form we assume a stellar age of 0.1 Myr when embryos have a mass of 
0.025 $M_\oplus$, 0.5 Myr when the embryos have a mass of 0.05 $M_\oplus$ and a disc age of 1 Myr when the embryos have a mass of 0.08 
$M_\oplus$ \citep{B16}.} The ratio of the total mass in the equal-mass embryos versus planetesimals is either 1:1, 3:1 or 7:1. {\ 
The planetesimals exert dynamical friction on the planets. The different total mass ratios between the embryos and the planetesimals 
were chosen to study how the final orbital eccentricities and inclinations of the planets depend on this initial mass ratio.}\\

The system consisting of gas giants, planetary embryos and planetesimals is simulated with the symplectic integrator package 
SyMBA \citep{Dll98} for 150~Myr using a time step of 7.3~days. The migration of the gas giants was mimicked through fictitious forces 
\citep{W11}. The gas disc model is based on that of \citet{B14}, with an initial surface density of 2272~g~cm$^{-2}$. {\ The disc's 
surface density at a fixed location decreases with time according to $\Sigma \propto t^{-7/5}$ \citep{H98}.} After 5 Myr, {\ when 
the stellar accretion rate drops below $10^{-10}$ $M_\odot$~yr$^{-1}$}, we artificially photoevaporated the disc away {\ 
exponentially with an e-folding time of} 100 kyr \citep{B14}. The planetary embryos experienced type I migration and tidal damping of 
their eccentricities and inclinations from the gas disc \citep{TW04}. The planetesimals experience gas drag, for which we assumed each 
planetesimal had a radius of 50 km. SyMBA treats collisions between bodies as perfect mergers, preserving their density. This works 
well in most circumstances, but given that {\ all the terrestrial planets have different densities from each other and from most 
meteorites} we modified SyMBA and implemented a mass-radius relationship that fits well through Mars, Venus and Earth \citep{S07}. 
This relation is applied after each collision to ensure that the final planets have radii comparable to the current terrestrial 
planets. {\ Initially all planetary embryos and planetesimals have a bulk density of 3 g cm$^{-3}$.} \\

The correlated isotope data for the various planets and meteorites, and their sources, are listed in the Supplementary Material.\\

{\ The isotopic, and therefore compositional, differences between Earth and Mars can possibly be quantified by tracking the initial 
heliocentric distance of material that will be incorporated into the planets. Assuming that the disc follows a simple relation between 
isotopic ratios and distance to the Sun it is possible to crudely quantify a bulk composition for all of the planets. However there is 
no well-established method for doing so. For example, it has been suggested that the primordial disc was homogeneous in silicon 
isotopes \citep{P13b}, while there appears to be a gradient in $^{54}$Cr and $^{50}$Ti \citep{Y10}, implying that several 
relationships ought to be used for different systems. A parsimonious approach is to assume that solar system material is increasingly 
oxidised with heliocentric distance, with very reduced material residing closer than 1.2~AU and more oxidised material farther out 
\citep{R15}. This can then be tied to the major meteorite groups. The most recent attempt, which is based on the Grand Tack 
model, suggested that enstatite chondrites resided between 1~AU to 2~AU, ordinary between 2~AU and 3~AU and the carbonaceous 
chondrites come from beyond 6~AU \citep{FGK17}.\\

Here we opt to use the composition of the asteroid belt to partition the different meteorite groups \citep{DC14}: enstatites 
originate within 1.5~AU, the ordinary chondrites resided between 1.5~AU and 3~AU and carbonaceous chondrites farther than 3~AU. In the 
next section we record the initial positions of embryos and planetesimals and how they assemble into the final planets. Since their 
initial position is a proxy for isotopic composition, we are able to characterise pathways for the Mars analogue that are consistent 
with its compositional affinity for ordinary chondrites, and thus quantify any compositional differences with Earth.}

\section{Results}
\begin{figure}[ht!]
\includegraphics[width=90mm]{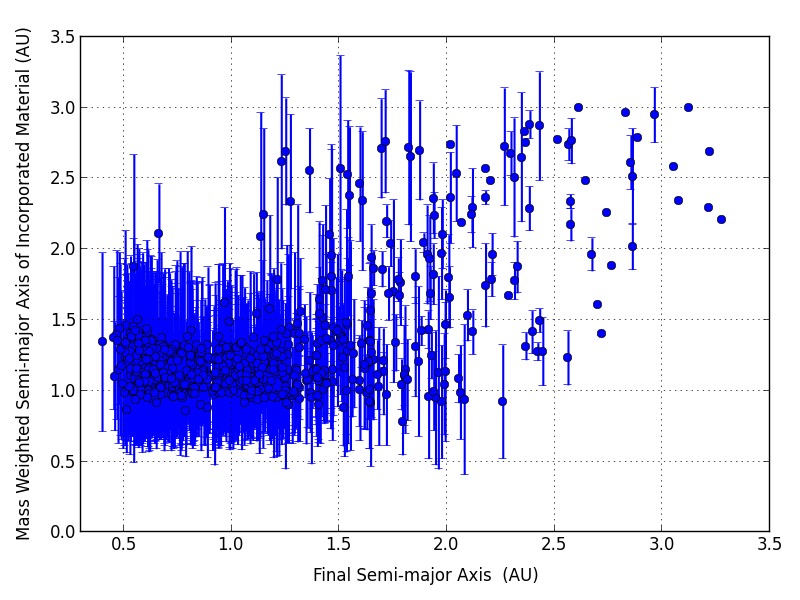}
\caption{Mass-weighted initial semi-major axis of material incorporated into the planets as a function of their final semi-major axis. 
The Grand Tack clears the disc beyond 1~AU and mixes in some material from farther out, so that there are many planets that only 
sample the inner disc.}
\label{fig:sumgt15}
\end{figure}

We present the summary of our Grand Tack simulations with a tack at 1.5~AU in Fig.~\ref{fig:sumgt15}, which depicts the mass-weighted 
mean initial semi-major axis of material incorporated into the planets versus the final semi-major axis of all the resulting planets 
from the simulations; this encompasses the contribution from both embryos and planetesimals. The error bars are weighted 
standard deviations. It is clear that the majority of planets end up between 0.5~AU to 1.3~AU and they sample the region of the 
protoplanetary disc between 0.7~AU and 1.7~AU. This outcome is not surprising, because Jupiter's incursion empties out the disc beyond 
approximately 1.2~AU while also pushing some material inwards. Planets that are situated further either escaped ejection or were 
tossed outward from the annulus inside of 1.2~AU. Planets situated near 1.5~AU tend to have a mass-weighted mean initial semi-major 
axis of 1.2~AU, but a few cases have higher values, beyond 1.5~AU. {\ Following \citet{B16}, we define a Mars analogue having a 
final semi-major axis between 1.3~AU and 1.7~AU. We also require that its mass $0.05 M_\oplus < m_p < 0.15 M_\oplus$.} \\

\begin{figure*}[ht!]
\includegraphics[width=200mm]{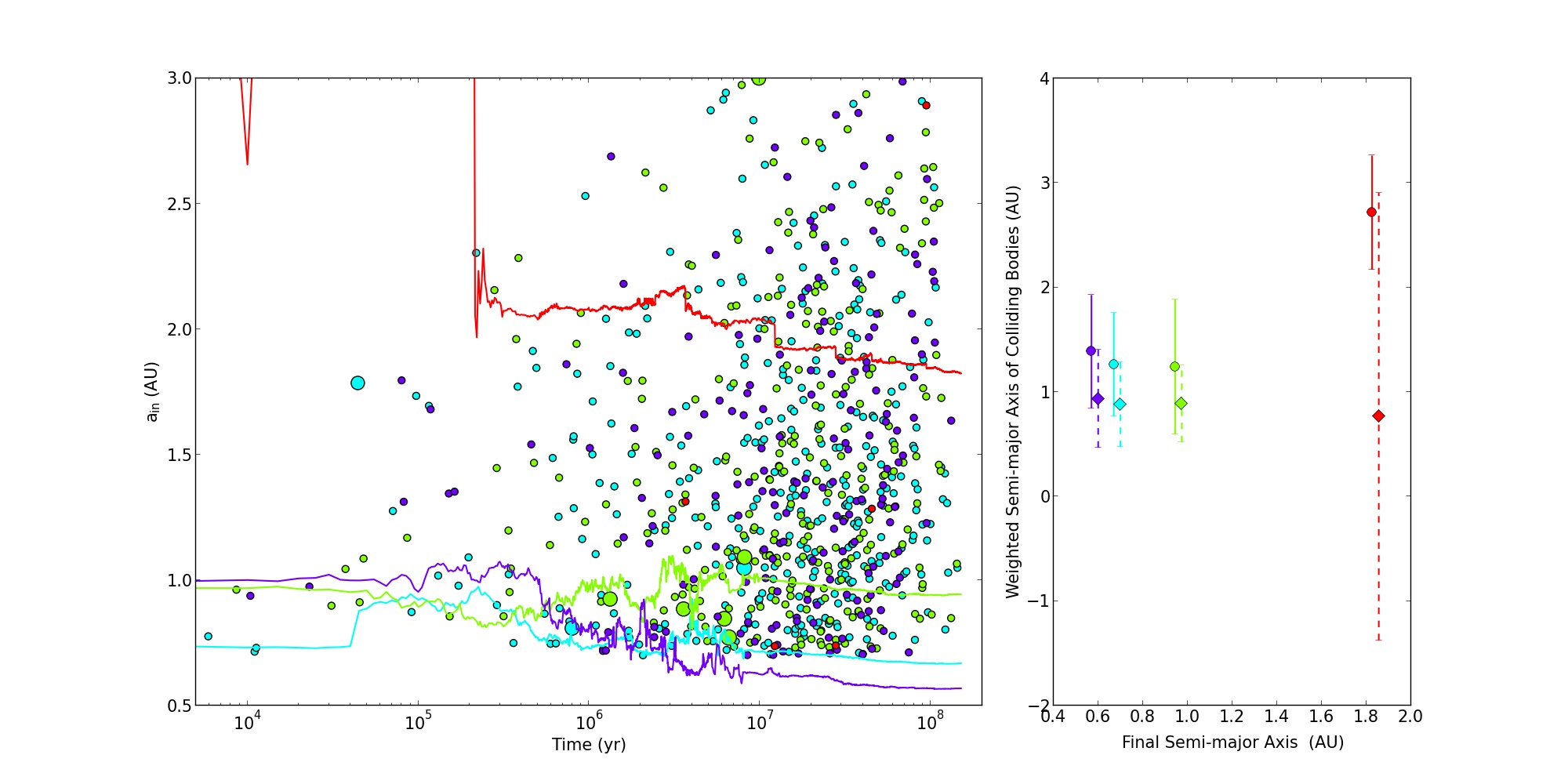}
\caption{Dynamical evolution of an extreme case that yields a Mars analogue that is compositionally different from the Earth and Venus 
because it samples a distinctly different region of the protoplanetary disc. The left panel indicates the evolution of the semi-major 
axis of the four terrestrial planet analogues and the coloured dots indicate the initial semi-major axis and time of bodies that 
collide with the terrestrial planet analogues. The size of the dots is indicative of their mass. The Mars analogue gets scattered 
first outwards and then inwards by Jupiter and slowly migrates towards 1.8~AU through dynamical interaction with the remaining 
planetesimals. The right panel indicates the mass-weighted mean initial semi-major axis of material that is incorporated into the four 
terrestrial planet analogues. The error bars depict the total range of material incorporated into the final planets in the Grand Tack 
model while the dotted error bars are the same if we suppose the planets originated from the same narrow region between 0.7~AU and 
1~AU 
as in the annulus model.}
\label{fig:evogood}
\end{figure*}
In Fig.~\ref{fig:evogood} our extreme case {\ yields a Mars analogue that is compositionally very different from the Earth}. The 
final planetary system has a Mercury analogue of 0.08~$M_\oplus$ at 0.55~AU, a Venus analogue of 0.97~$M_\oplus$ at 0.69~AU, an Earth 
analogue of 0.52~$M_\oplus$ at 1.12~AU and a Mars analogue of 0.12~$M_\oplus$ at 1.79~AU. The left panel depicts the evolution of the 
semi-major axes of the terrestrial planet analogues, with the coloured dots indicating the time and initial semi-major axis of embryos 
and planetesimals that collided with the planets. The Mars analogue started as an embryo beyond 3~AU, and apart from a very brief 
excursion to the outer Solar System due to encounters with Jupiter, landed near its current location. {\ We add that it is more 
typical for a distant Mars analogue to be pushed inwards through capture in the 2:1 resonance with Jupiter rather than through 
repeated scattering.}\\

The right panel plots the final semi-major axis versus the mass-weighted mean initial semi-major axis of the terrestrial planet 
analogues. The solid error bars depict the total range of material incorporated into the final planets in the Grand Tack model while 
the dotted error bars are the same if we suppose the planets originated from the same narrow region between 0.7~AU and 1~AU as in the 
annulus model. In this simulation all material incorporated into the Mars analogue comes from beyond 2~AU in the Grand Tack case, 
while in the annulus model the material accreted by all planets would have originated inside of 1~AU (by definition, because initially 
there is no solid material beyond 1~AU). On the other hand, material accreted by the Earth and Venus analogues mostly samples the 
inner region of the disc for both models, suggesting a very distinct composition from the Mars analogue.\\

\begin{figure*}[ht!]
\includegraphics[width=200mm]{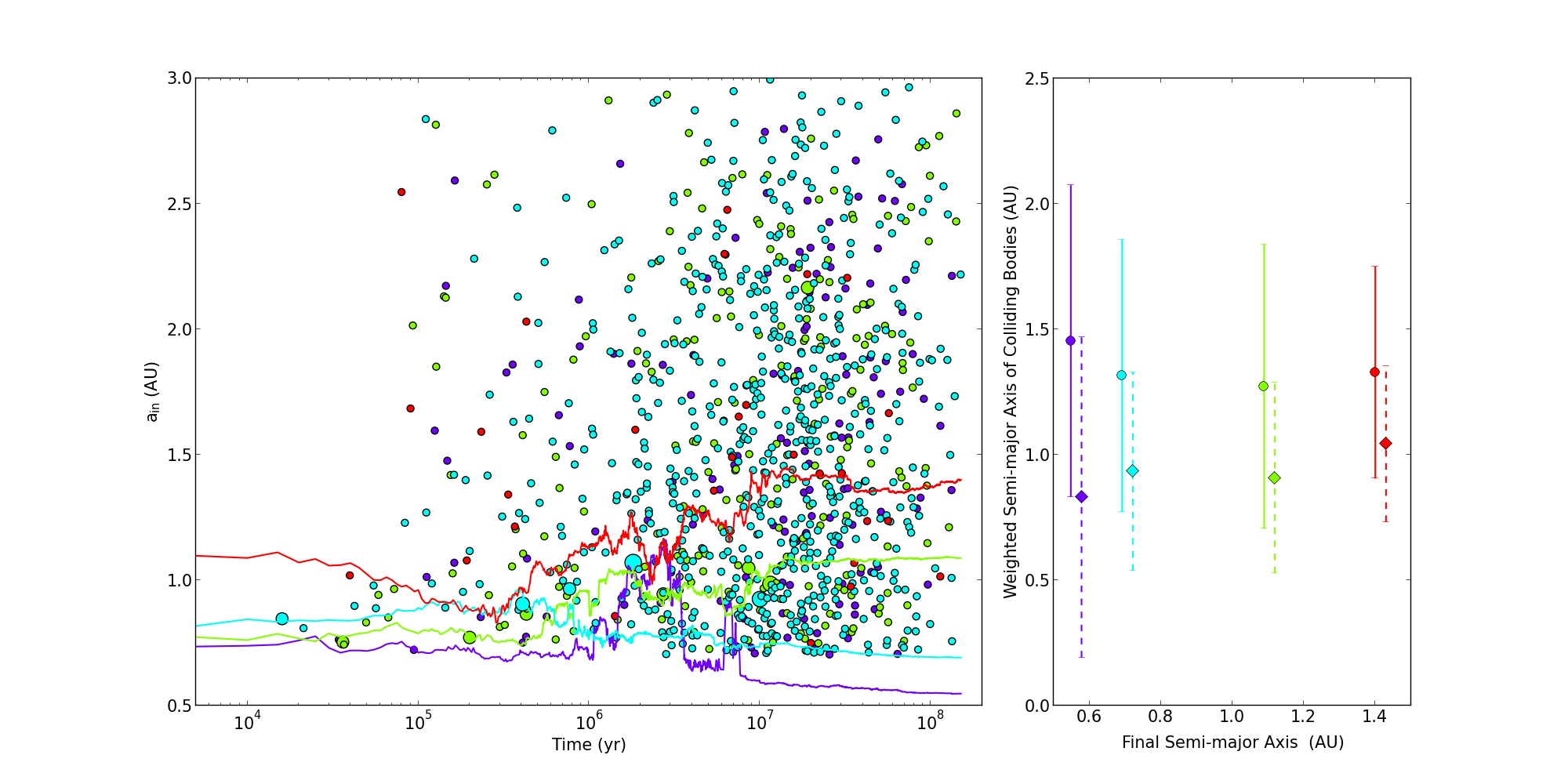}
\caption{Same as Fig.~\ref{fig:evogood} but now we show a case where the composition of the Mars analogue is nearly identical to that 
of the Earth analogue because they sample the same region of the protoplanetary disc.}
\label{fig:evobad}
\end{figure*}
{\ We now compare our results with those of \citet{FB16}.} As mentioned above, \citet{FB16} attempted to reproduce the composition 
of the terrestrial planets by sampling different source materials, reflected in different meteorite groups, from the protoplanetary 
nebula in a Monte Carlo approach. {\ \citet{FB16} conclude that Earth and Mars have a very similar bulk composition, that is both 
planets shared the same (mixed) source material. Such an outcome is only possible if both worlds accreted the same source material 
i.e. 
their feeding zones greatly overlapped. Dynamically the two planets can only have a nearly identical composition if they} formed close 
to each other in a common narrow annulus of the protoplanetary disc, with Mars subsequently scattered out of the annulus to its 
present position, which prevented further accretion.\\

We present the dynamical evolution of such a case in Fig.~\ref{fig:evobad}, where the final planets are a Mercury analogue of 
0.17~$M_\oplus$ at 0.55~AU, a Venus analogue of 1.13~$M_\oplus$ at 0.69~AU, an Earth analogue of 0.59~$M_\oplus$ at 1.09~AU and a Mars 
analogue of 0.06~$M_\oplus$ at 1.40~AU. Although this type of evolution where Mars originates within 1~AU and is subsequently 
scattered outwards is the most common, it is inconsistent with the compositional difference between Earth and Mars, not least the 
increased ordinary chondrite component in Mars compared with Earth. {\ Thus, for Earth and Mars to have a very different 
composition} we argue that {\ Mars must have predominantly assembled from material beyond 1.5~AU, of which Earth accreted 
relatively little}.\\

\begin{figure}[ht!]
\includegraphics[width=90mm]{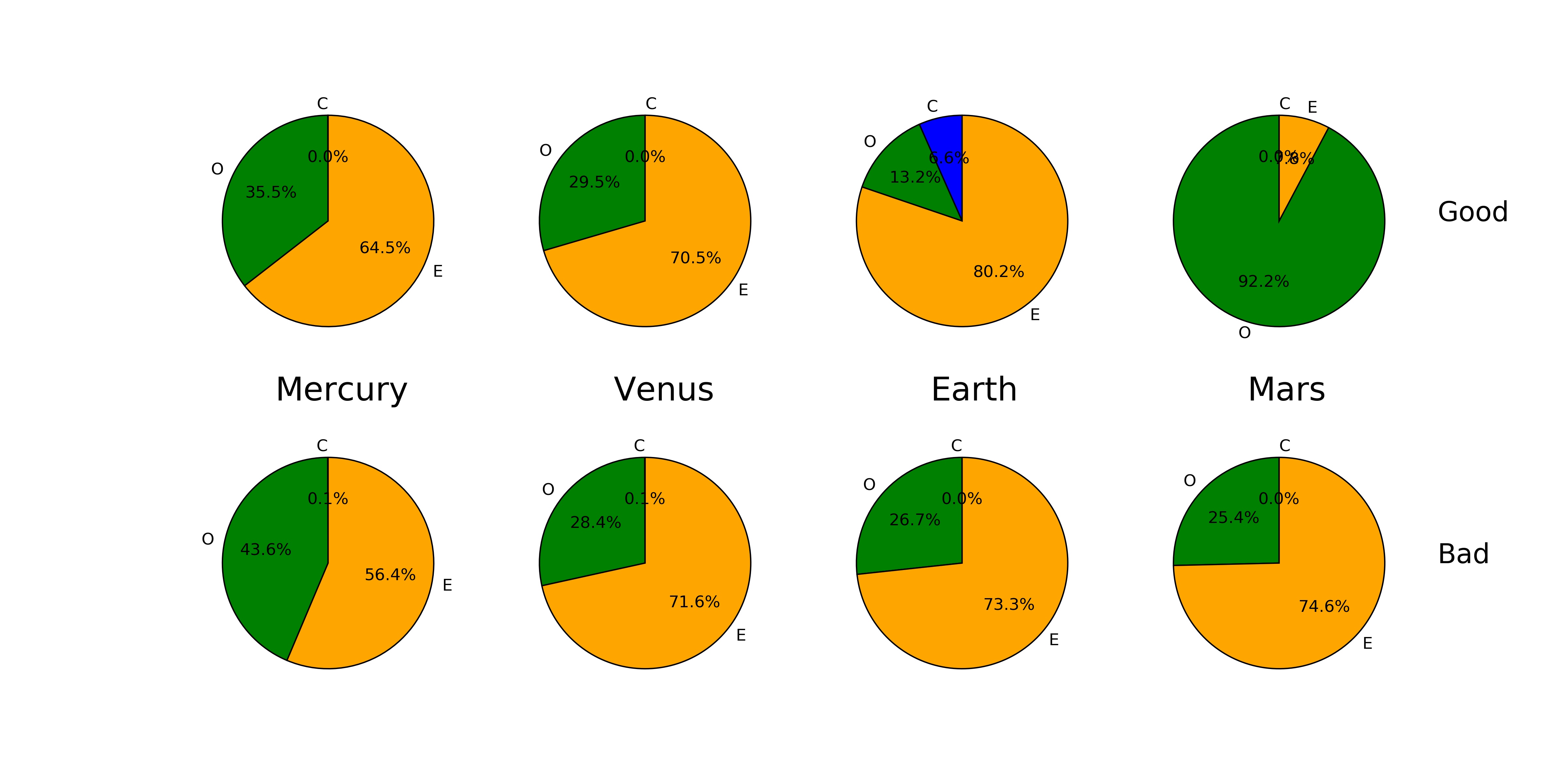}
\caption{Pie chart of the composition of the terrestrial planets for the successful simulation, corresponding to  
Fig.~\ref{fig:evogood} (top), and the unsuccessful simulation in Fig.~\ref{fig:evobad} (bottom). Labels indicate which chart 
corresponds to what planet analogue. Enstatite chondrite is in orange, ordinary chondrite in green and carbonaceous in blue. In the 
top case the Earth analogue accreted some carbonacous chondrite material. Note that the successful case has a composition for Mars 
that 
is strikingly different than that of the other three planets.}
\label{fig:pies}
\end{figure}

\begin{figure}[ht!]
\includegraphics[width=90mm]{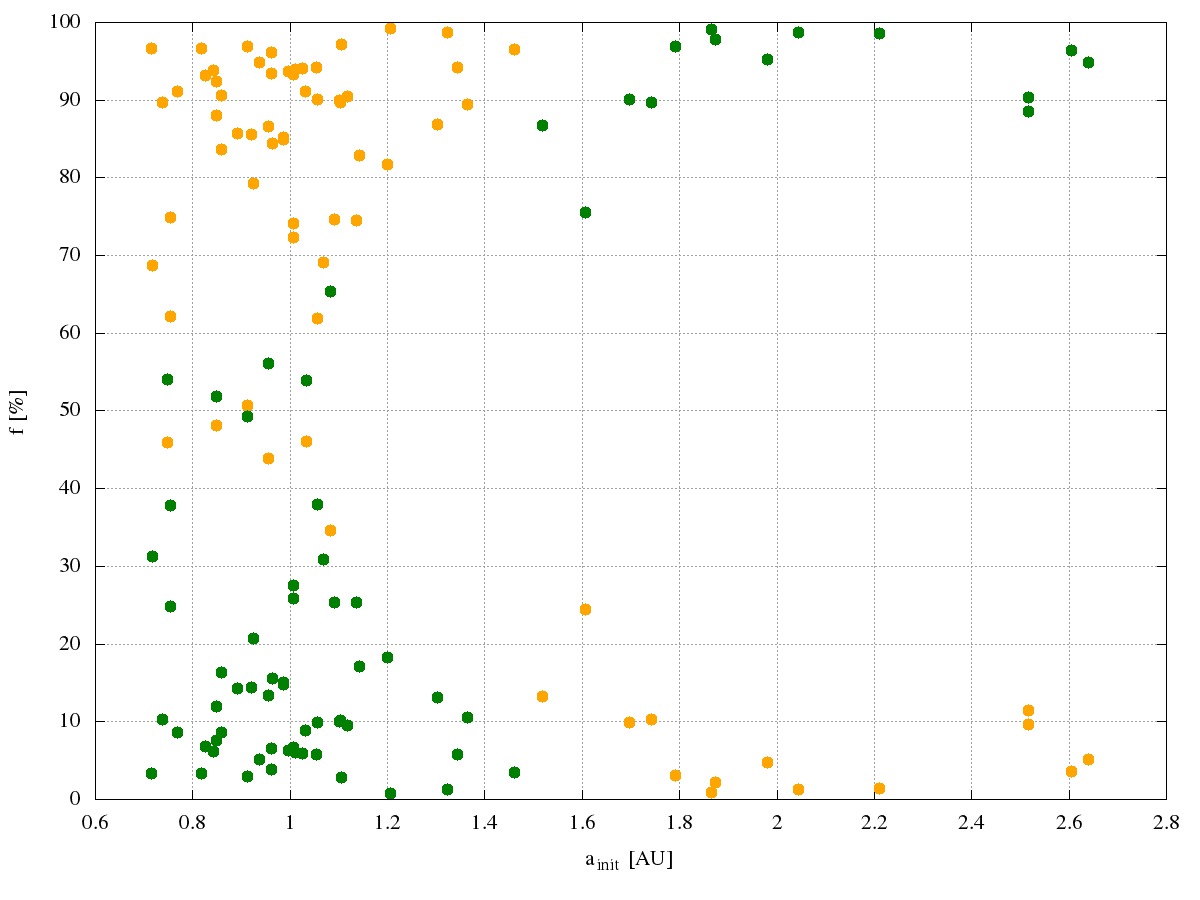}
\caption{Scatter plot of the percentage of enstatite (orange) and ordinary chondrite (green) compositional make-up of all Mars 
analogues. Two dots are plotted for each planet, with the combined values adding up to 100\%.}
\label{fig:ecoc}
\end{figure}

{\ Under the assumption of the compositional gradient in the disc presented earlier, we find that on average the Venus, Earth and 
Mars analogues are composed of 82\% $\pm$ 10\%, 87\% $\pm$ 8\% and 69\% $\pm$ 30\% enstatite chondrite and 18\% $\pm$ 10\%, 13\% $\pm$ 
7\% and 31\% $\pm$ 30\% ordinary chondrite. The fraction of carbonaceous chondrite is insignificant, which is most likely an artefact 
of our initial conditions since the solid disc only extended to 3~AU \citep{B16}. These values are in agreement within uncertainties 
with earlier studies \citep{S99,TD14,D14}. In Fig.~\ref{fig:pies} we show the fraction of these materials in each planet for the two 
simulations discussed in Fig.~\ref{fig:evogood} and Fig.~\ref{fig:evobad}. In the top panel, corresponding to the evolution where Mars 
formed far, the inner three planets are dominated by enstatite chondrite material while the Mars analogue consists mostly of ordinary 
chondrite material (which is consistent with Fig.~\ref{fig:isotopes}). In the bottom panel, where Mars formed closer in, all four 
planets have a very similar composition. This compositional homogeneity is in agreement with \citet{FB16} but is simultaneously 
inconsistent with the isotopes presented in Fig.~\ref{fig:isotopes}.\\

The large uncertainties in the compositional makeup of Mars imply that there is a great variety, with some cases consisting of mostly 
enstatite chondrite and others that are mostly ordinary chondrite. In Fig.~\ref{fig:ecoc} we plot the percentage of the enstatite and 
ordinary chondrite components for all Mars analogues versus their initial semi-major axis. For each planet, two dots are plotted that 
combined add up to 100\%. The transition from mostly enstatite to ordinary chondrite composition at 1.5~AU is by choice. What is 
important, however, is that Mars analogues that started closer than 1.2~AU have a fully mixed composition, which is expected from 
Fig.~\ref{fig:sumgt15}, while those that formed farther out are predominantly ordinary (beyond 1.5~AU) or enstatite (between 1.2~AU 
and 1.5~AU).}\\

\begin{figure}[ht!]
\includegraphics[width=90mm]{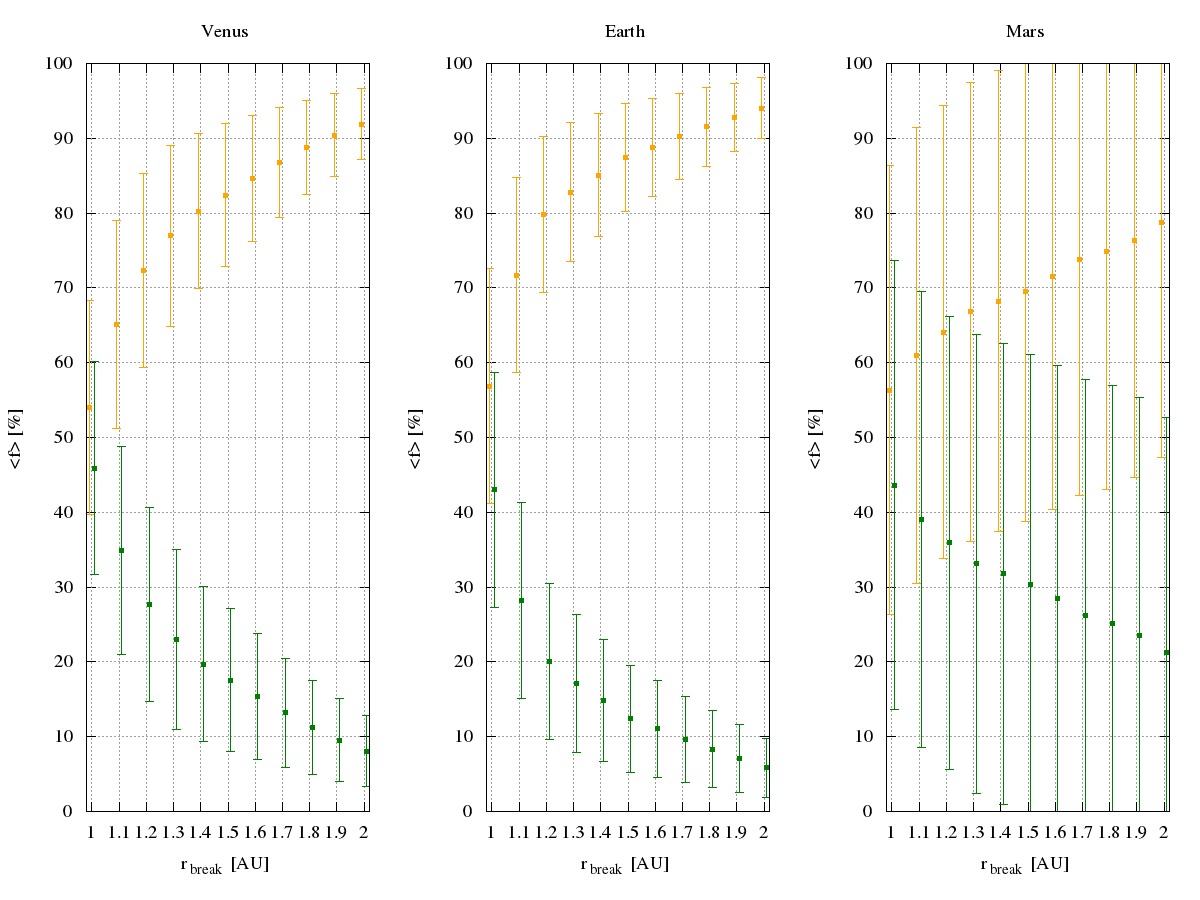}
\caption{Scatter plot of the average percentage contribution of enstatite (orange) and ordinary chondrite (green) to the bulk 
compositions of Venus, Earth and Mars. Data are plotted as a function of the distance where the disc changes its composition from one 
meteorite class to another.}
\label{fig:comps}
\end{figure}

{\ Fig.~\ref{fig:comps} reports our computed averages for the percentage contribution of enstatite (orange) and ordinary (green) 
chondrite material to the bulk chemistries of Venus, Earth and Mars. The analysis is as a function of the distance from the Sun where 
the dominant composition of the disc changes from one category of meteorite class to another. The error bars shown are 1-$\sigma$ 
values. Results show that a changeover distance of 1.7~AU best matches Earth's composition of approximately 90\% enstatite and 10\% 
ordinary chondrite. This location, however, does not match Mars' 45\% enstatite and 55\% ordinary chondrite composition which 
underscores the notion that -- at least for Mars -- the underlying dynamics of its formation matter in its ultimate compositional 
make-up, as revealed in Figs.~\ref{fig:evogood} and~\ref{fig:ecoc}.}.\\

\begin{figure}[ht!]
\includegraphics[width=90mm]{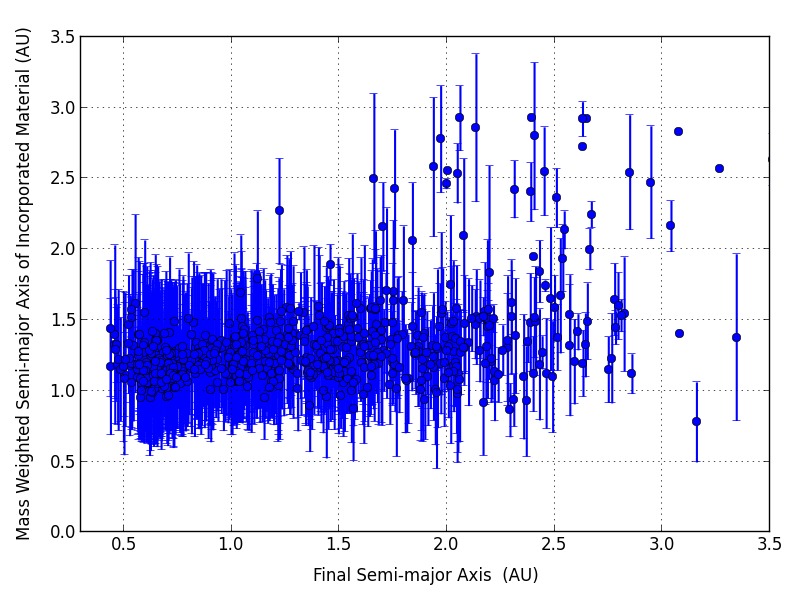}
\caption{Mass-weighted initial semi-major axis of material incorporated into the planets as a function of their final semi-major axis 
for the set of Grand Tack simulations with the tack at 2~AU. Note that the majority of planets still only sample the inner part of the 
disc and that this region now extends to beyond 2~AU.}
\label{fig:sumgt20}
\end{figure}

In all the various simulations performed with a tack at 1.5~AU we form a cumulative output of 635 terrestrial planets. Of these 
different simulations and their various outputs, a total of {\ ten} are Mars analogues with {\ an initial semi-major 
axis greater than 1.5~AU}. This is an effective production probability of 1.6\% $\pm$ 0.5\% accounting for Poisson statistics, and as 
such makes this a very unlikely avenue for Mars' formation. {\ The region of the disc beyond 1.5~AU is primarily composed of 
ordinary chondrite materials, which is what is likely the major constituent of Mars' bulk composition.} The total fraction of Mars 
analogues is 8.9\% $\pm$ 1.1\%. Should the tack of Jupiter have occurred at 2~AU, the mass-weighted mean initial semi-major 
axis of planetary material versus the final semi-major axis of all the planets for the solar system would look like depicted in 
Fig.~\ref{fig:sumgt20}. In this condition, there are a total of 685 planets in our cumulative output, of which just {\ eight} are 
Mars analogues with {\ an initial semi-major axis greater than 1.5~AU}, corresponding to a probability of {\ 1.1\% $\pm$ 
0.4\%.} This is {\ comparable to} the case that the tack occurred at 1.5~AU. {\ The final compositions of the planets are 
identical within uncertainties to the case with a tack at 1.5~AU.} \\

\section{Discussion and Conclusions}
We previously concluded that the orbital distribution of the terrestrial planets was better reproduced with a tack location of 
Jupiter at 2~AU rather than at 1.5~AU \citep{B16}. The results of the present study, on the other hand, {\ indicate no preference}. 
Further study on the tack location is warranted.\\

A more distant formation of Mars has obvious consequences beyond its bulk composition, to its proclivity to have established a 
biosphere. Although a topic of active debate, Mars could have retained a higher fraction of its volatile inventory than Earth 
\citep{K14} in the absence of collisional erosion. There is no doubt that Mars was bombarded by asteroids and comets 
(e.g. \cite{AM16}), but appears to have escaped the kind of devastating late giant impacts one the scale of which generated Earth's 
Moon. This separate history is consistent with analysis of martian meteorites which show relatively enhanced complements of 
moderately-volatile elements compared to Earth (e.g. \cite{AG13,W15}), {\ which is most probably a result of its higher fraction of 
ordinary chondrite material (see Supplementary materials).} Delivery and retention of a hydrosphere are deemed important for a 
biosphere, but a consistently more distant location for Mars also calls for an historically much lower solar flux. {\ Unless, as our 
model shows, an intrinsically volatile-rich Mars possessed a strong and sustainable greenhouse atmosphere, its average surface 
temperature was unremittingly below 0$^\circ$~C \citep{F13}}. Such a cold surface environment would have been regularly affected by 
early impact bombardments that both restarted a moribund hydrological cycle, and provided a haven for possible early life in the 
martian crust \citep{AM16}.\\

Our study also predicts that Venus' bulk composition, including its oxygen isotopes, should be comparable to that of the Earth-Moon 
system because our dynamical analyses show they share the same building blocks. Indeed, in our simulations we find that the 
Venus and Earth analogues always share the same material, whereas the same cannot be said for Earth and Mars. Recent 
ground-based IR observations of the Venusian atmosphere yield deviations in the three-oxygen isotope system that overlap with the 
Earth-Moon fractionation line within errors. This is the expected result if the material that gave rise to Venus was mixed with that 
which built the Earth \citep{I15}.\\

{\ Given our predictions for the composition of the terrestrial planets and our simplified inverse modelling of the composition of 
the primordial disc, a pertinent follow-up study would consist of assuming a variety of locations of the transition from enstatite 
to ordinary chondrite materials and coupling this to our dynamical simulations in a Monte Carlo fashion.}

\section{Acknowledgements}
We are grateful to A. Morbidelli, N. Dauphas and C. Burkhardt for valuable feedback during the early stages of this work. 
We also thank S. Charnoz and an anonymous reviewer for constructive comments that substantially improved this manuscript, and to 
F. Moynier for editorial assistance and for useful insights. RB is grateful for financial support from the Daiwa Anglo-Japanese 
Foundation and JSPS KAKENHI (16K17662). RB and SJM acknowledge The John Templeton Foundation - FfAME Origins program in the support of 
CRiO. SJM is grateful for support by the NASA Exobiology Program (NNH14ZDA001N-EXO).

\end{document}